# Trion confinement in monolayer MoSe$_2$ by carbon nanotube local gating

Takato Hotta[1], Haruna Nakajima[1], Shohei Chiashi[2], Taiki Inoue[2], Shigeo Maruyama[2], Kenji Watanabe[4], Takashi Taniguchi[3], and Ryo Kitaura[1,3,*]

[1]*Department of Chemistry, Nagoya University, Nagoya, Aichi 464-8602 Japan*

[2]*Department of Mechanical Engineering, The University of Tokyo, 7-3-1 Hongo, Bunkyo-ku, Tokyo, 113-8656 JAPAN*

[3]*International Center for Materials Nanoarchitectonics, National Institute for Materials Science, 1-1 Namiki, Tsukuba 305-0044, Japan*

[4]*Research Center for Functional Materials, National Institute for Materials Science, 1-1 Namiki, Tsukuba 305-0044, Japan*

Corresponding Author: R. Kitaura, KITAURA.Ryo@nims.go.jp, r.kitaura@nagoya-u.jp

Abstract

We have successfully confined trions into a one-dimensional restricted space of a MoSe$_2$ device with CNT gate electrodes. The dry transfer process, including deterministic dry transfer of aligned CNTs, has led to an hBN-encapsulated MoSe$_2$ device with CNT back gate electrodes. In contrast to a location without CNT gate electrodes, applying voltage via CNT gate electrodes significantly alters PL spectra at a location with CNT gate electrodes. PL imaging has revealed that image contrast from trions is linear along the CNT electrode underneath, consistent with 1D confinement of trions in response to the CNT local gating. The confinement width obtained from the PL image is $5.5 \times 10^2$ nm, consistent with nanoscale 1D confined trions with the diffraction limit broadening. This work has demonstrated electrical control of excitonic states at the nanoscale, leading to novel optoelectronic properties and exciton devices in the future.

Introduction

Electrical confinement and manipulation of carriers are the basis for quantum optoelectronic devices. For example, in the high-electron-mobility transistor (HEMT), two-dimensional (2D) electron gas confined at the interface between n-doped AlGaAs and GaAs can be electrically controlled, working as field-effect transistors with high carrier motilities.[1, 2] Also, electrons confined in multiple 2D quantum wells can respond to infrared light irradiation under bias voltage, leading to high responsivity infrared photodetectors.[3] In addition to the confinement into 2D space, confinement of electrons into zero-dimensional (0D) space, i.e., semiconductor quantum dots, also allows the development of quantum devices, where coherent manipulation of electron spins has been successfully demonstrated.[4-6]

In addition to the quantum device applications, the confinement of electrons and holes into low-dimensional space provides a versatile playground to explore fundamental optical physics. For example, photogenerated carriers confined in 2D semiconductors, such as transition metal dichalcogenides (TMDs), form hydrogenic bound states, called excitons, even at room temperature due to the enormous exciton binding energy arising from 2D structures.[7] In addition to excitons, electrons and holes confined in 2D TMDs lead to various excitonic complexes, such as trions, biexcitons, and charged biexcitons with the spin and valley-degree-of-freedom.[8-13] Furthermore, recent studies on TMD-based van der Waals (vdW) stackings have shown that TMDs can also host excitons confined in 0D space because of 0D confinement potential (the moiré potential) originating from interlayer vdW interaction; these 0D excitons lead to the realization of quantum emitters embedded in 2D TMDs.[14-17] These examples have clearly shown that controlled confinement and manipulation of excitonic excitations can further contribute to fundamental optical physics, which triggers the development of novel quantum optoelectronic devices.

Controlled electrical confinement of excitonic states, however, has been challenging. An exciton composed of an electron and a hole is a neutral quasiparticle, insensitive to the electric field for electrical confinement. In addition, excitons are usually not stable at room temperature, demanding the cryogenic temperature to manipulate excitons.[18] A recent study has shown that trions (charged excitons) in a 2D TMD can be confined into sub-micrometer 0 and 1D regions by applying the electric field via split-gate electrodes.[19] Although this seminal work has successfully shown the electrical control of trions, the controllable confinement of excitonic excitations into nanoscale space is still challenging.

To realize controlled nanoscale confinement of excitonic states, we have focused on trions confinement with a carbon nanotube (CNT) as a gate electrode. CNTs are natural ultimately-thin conductors with 1 ~2 nm in diameter, which is beyond the limit of the current microfabrication technique.[20] We can form 1D confinement potential in 2D semiconducting TMDs by applying the local electric field through the CNT confinement gate adjacent to 2D TMDs. For this purpose, we have developed a method for the deterministic transfer of CNTs with high position and angle accuracy. We used aligned CNTs grown on a quartz substrate to place CNT confinement gate electrodes; the polymer-based stamp technique can directly pick aligned CNTs up and put them onto a $SiO_2$/Si substrate without losing orientational alignment.[21] We found that a CNT confinement gate electrode can confine trions into 1D space in an hBN-encapsulated monolayer $MoSe_2$ with the CNT confinement gate placed at the bottom of the stacked structure. The photoluminescence (PL) imaging performed at 10 K has demonstrated that the PL intensity arising from trions appears along the CNT confinement gate electrode with a width of $5.5 \times 10^2$ nm. Considering the broadening from the diffraction limit, the observed 1D PL contrast is consistent with the width of charge density induced by the CNT gate electrode.

Results and discussion

The aligned CNTs were grown by the alcohol catalytic chemical vapor deposition (CVD) method on the R-cut quartz substrates [22, 23]. Figure 1(a) shows a scanning electron microscope (SEM) image of the aligned CNTs on a quartz substrate, where white vertical and horizontal lines correspond to the grown CNTs and deposited metals for catalyst, respectively. The highly-aligned and isolated CNTs, whose typical length reaches 100 μm, are essential for fabricating devices with CNT gate electrodes. The aligned CNTs are grown from the catalyst lines composed of metal nanoparticles and CNT assemblies, and it is easy to pick the whole CNTs-catalyst line structure with a polymer stamp.

Figure 1(b) shows a schematic representation of the CNT placing process. For the deterministic placement of CNTs, applying the stamp-based dry transfer technique is essential. However, CNTs grown on quartz substrates are strongly attached to the surface, making applying the dry transfer method challenging. For this reason, we firstly transferred CNTs grown on the quartz to $SiO_2$/Si substrate by the wet transfer method, where an aqueous solution of KOH facilitates the detachment of the CNTs from quartz substrates. Figure 2(a) shows optical microscope images of CNTs on a $SiO_2$/Si substrate before and after a pickup process; polycarbonate (PC)/polydimethylpolysiloxane (PDMS) stamps (a typical image of a stamp is shown in Fig. 1(c)) were used in this work. As seen in these images, the linear contrasts from dense CNTs grown along the growth catalyst disappear after the pickup process, indicating that a PC/PDMS stamp can pick them up easily once CNTs are placed on $SiO_2$/Si. Figures 2(b) show a typical AFM image of CNTs placed onto a $SiO_2$/Si substrate (100 nm $SiO_2$). As can be seen from the Figure, the linear CNT alignment is maintained throughout the transfer process. Since this method can retain the CNT alignment, two consecutive transfers can yield a rectangular lattice of CNTs (Fig. 2(c) top and bottom).

We have fabricated an hBN-encapsulated monolayer MoSe$_2$ structure with a bottom CNT gate electrode based on the CNT placement technique. First, we placed CNTs on a SiO$_2$/Si substrate by the dry transfer process. And then, a few-layer graphene flake was picked up with an hBN flake on a PC/PDMS stamp. Subsequently, the MoSe$_2$ flake and another hBN flake (the bottom layer) were picked up consecutively, and the resulting hBN/MoSe$_2$/hBN was put onto the CNTs/SiO$_2$/Si. After removing the PC film by dipping it into dichloromethane for half a day, the electron beam lithography and reactive ion etching techniques were used to make electrical contacts and remove extra CNTs on the substrate. Figures 3(a) and (b) show a schematic and an optical image of the fabricated MoSe$_2$ device. The AFM image corresponding to the red box in Fig. 3(b) shows the existence of CNT-bottom gates under the hBN/MoSe$_2$/hBN.

PL spectra of hBN/MoSe$_2$/hBN/CNT were measured at 10 K with an excitation wavelength of 550 nm. Figure 4(a) shows a PL spectrum on a CNT gate electrode with $V_{CNT} = 0$; hereafter, the voltage applied through the CNT gates is denoted as $V_{CNT}$. The Figure shows that two peaks appear at around 1.64 and 1.61 eV, assigned to exciton and trion radiative recombinations. The trion peak, consistent with previous works, probably originates from free trions (not 1D confined trions) in monolayer MoSe$_2$.[24] Since the spot diameter of the excitation laser is much larger than the diameter of the CNTs, free trions should be produced in response to laser excitation. Figure 4(b) left shows the relation between $V_{CNT}$ and the trion emission measured at a location without CNT gate electrodes; we applied a Si back-gate voltage of −10 V to make all CNTs conductive. As seen in the Figure, the PL spectra are insensitive to $V_{CNT}$ throughout the measurements, $V_{CNT}.0 - 2$ V. On the other hand, PL spectra measured at a location with CNT gate electrodes significantly depend on $V_{CNT}$ (Fig. 4(b) right). PL intensity from trions increases, probably arising from carrier accumulation induced by $V_{CNT}$.

To visualize confined trions directly, we observed PL images of the device with $V_{CNT} = 2$ V; we tilted an 800 nm long-pass filter to filter out exciton emission and excitation light (broadband light λ < 550 nm) for selective observation of the trion emission. Additionally, a PL image taken with $V_{CNT} = 0$ V was subtracted to extract PL responding to the CNT gating selectively. As seen in Figure 5(a), the PL image shows a linear white contrast, which corresponds well to the location of the CNT gate electrode shown in Fig. 3(b); the AFM image is also shown for comparison. The correspondence between the contrast of the PL image and the position of the CNT gate electrode is consistent with the formation of trions via carrier injection by the CNT gate electrode. Figure 5(b) shows the PL intensity profile along the blue dashed line shown in the Figure. The profile can be fitted with a Gaussian function well, giving a standard deviation $\sigma_{obs}$ of $2.3 \times 10^2$ nm; the full width at half maximum (FWHM) of the PL intensity profile is ca. $5.5 \times 10^2$ nm.

Two possible factors in determining the width observed in the PL image are the broadening from the diffraction limit and the width of the confined trions. These two factors can cause broadening in PL images independently. For the broadening from the diffraction limit, we used the gaussian point spread function with $\sigma_{DL} = 0.21\lambda/NA$, where $\lambda$ and $NA$ correspond to the emission wavelength of trions (770 nm) and the numerical aperture of the objective lens (0.7), respectively.[25] For the broadening from the trion confinement, we used a simple model assuming a conductive plane and linear charge; the plane and linear charge correspond to monolayer $MoSe_2$ and CNT gate, respectively. In this case, the charge density induced on the plane (trions induced on $MoSe_2$) is described with the Lorentzian function, $\rho \propto d/(x^2 + d^2)$, where $d$ and $x$ correspond to the plane-linear charge distance and the position in the plane, respectively; the $d$ value, the distance between CNT and $MoSe_2$, corresponds to the thickness of hBN between $MoSe_2$ and CNT gate electrodes. Using the thickness of hBN ($d \sim 30$ nm), the FWHM of the trion confinement is estimated to be 32 nm. These two factors, Gaussian and Lorentzian broadening from the diffraction limit and the distribution of induced charges, can be represented as a convolution of the Lorentzian function with the Gaussian function. As seen in Fig. 5(b), a profile calculated based on the convolution reproduces the observed PL line profile well, clearly demonstrating that CNT local gating can lead to nanoscale confinement of excitonic states.

In conclusion, we have successfully confined trions into a 1D restricted space of a $MoSe_2$ device with CNT gate electrodes. The dry transfer process, including deterministic dry transfer of aligned CNTs, has led to an hBN-encapsulated $MoSe_2$ device with CNT back gate electrodes. In contrast to a location without CNT gate electrodes, applying voltage via CNT gate electrodes alters PL spectra at a location with CNT gate electrodes. PL imaging has revealed that image contrast from trions is linear along the CNT electrode underneath, consistent with 1D confinement of trions. The confinement width obtained from the PL image is $5.5 \times 10^2$ nm, which is consistent with 1D confined trions with the diffraction limit broadening. This work has demonstrated electrical control of excitonic excitation at the nanoscale, leading to novel optoelectronic properties and devices in the future.

This work was supported by JSPS KAKENHI Grant Numbers JP21K18930, JP20H02566, JP22H05458, JP20H05664, and JST CREST Grant Number JPMJCR16F3 and JPMJCR19H4, and JST PRESTO Grant Number JPMJPR20A2.

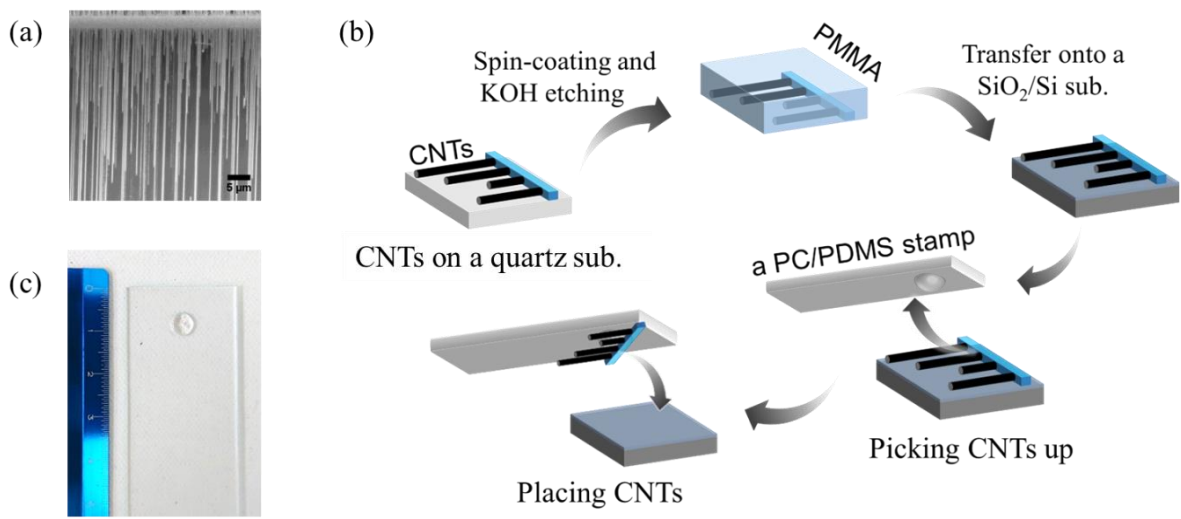

Figure. 1 (a) Typical SEM image of aligned CNTs on a quartz substrate. (b) Schematic representation of the CNT placing method. CNTs grown on a quartz substrate are transferred onto a $SiO_2$/Si by a wet transfer method with PMMA, and then the resulting CNTs on $SiO_2$/Si are picked up directly with a PC/PDMS stamp for deterministic placing. (c) Typical optical microscope image of a PC/PDMS stamp used in this work. A small stamp can be seen on a glass plate.

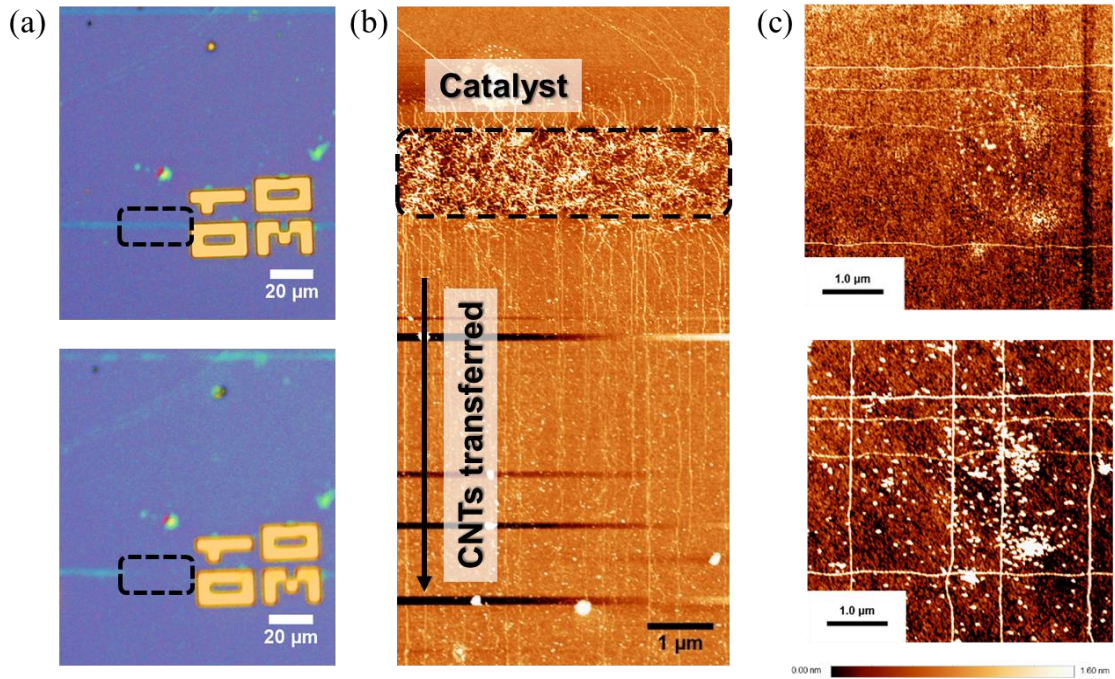

Figure 2 (a) Optical microscope images of CNTs on a SiO$_2$/Si substrate before and after picking CNTs up with a PC/PDMS stamp. The disappearance of a linear contrast from CNTs is seen in the black dotted frames. (b) Typical AFM image of CNTs placed on a SiO$_2$/Si substrate. Aligned CNTs perpendicular to the catalyst line is seen. The typical height of the aligned CNTs is 1~2 nm. (c) AFM images of aligned CNTs on a SiO$_2$/Si: (top) a horizontal alignment and (bottom) a grid-like alignment.

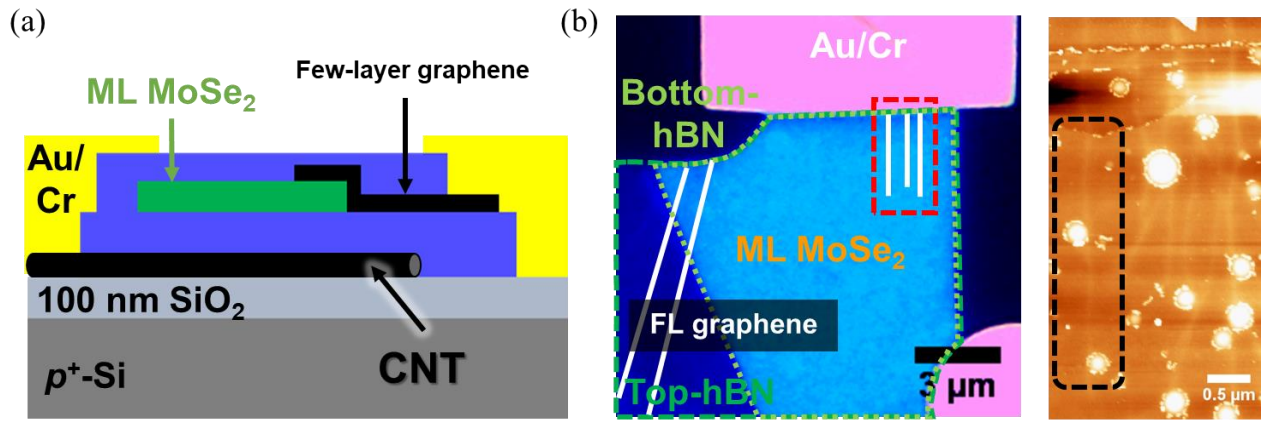

Figure 3 (a) Schematic image of the cross-section of the hBN/MoSe$_2$/hBN device with CNT back gate electrodes. A few-layer graphene flake was used to improve electrical contact with MoSe$_2$. (b) Optical image of a fabricated device and an AFM image of the device taken at the location shown as the red frame in the optical image. The black dotted frame corresponds to the location where a single CNT is seen; this place was investigated by PL imaging (see. Fig. 5(a)). Contrasts originating from impurities encapsulated between hBN and MoSe$_2$ are also seen.

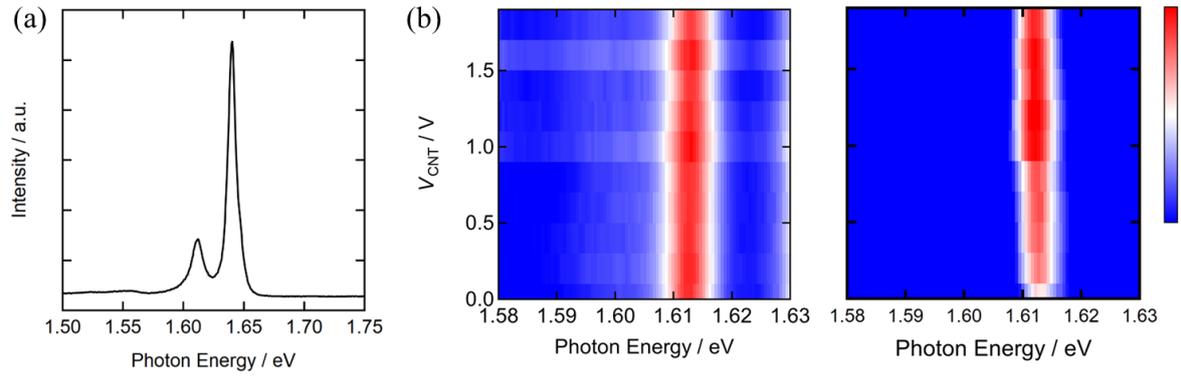

Figure 4 (a) PL spectrum measured at a location with a CNT gate and $V_{CNT}$ = 0. (b) 2D plots of $V_{CNT}$ dependence of PL spectra taken at locations without (left) and with (right) CNT back gate electrodes. The energy range of 1.58 – 1.63 eV is used to show the change of the trion peak clearly.

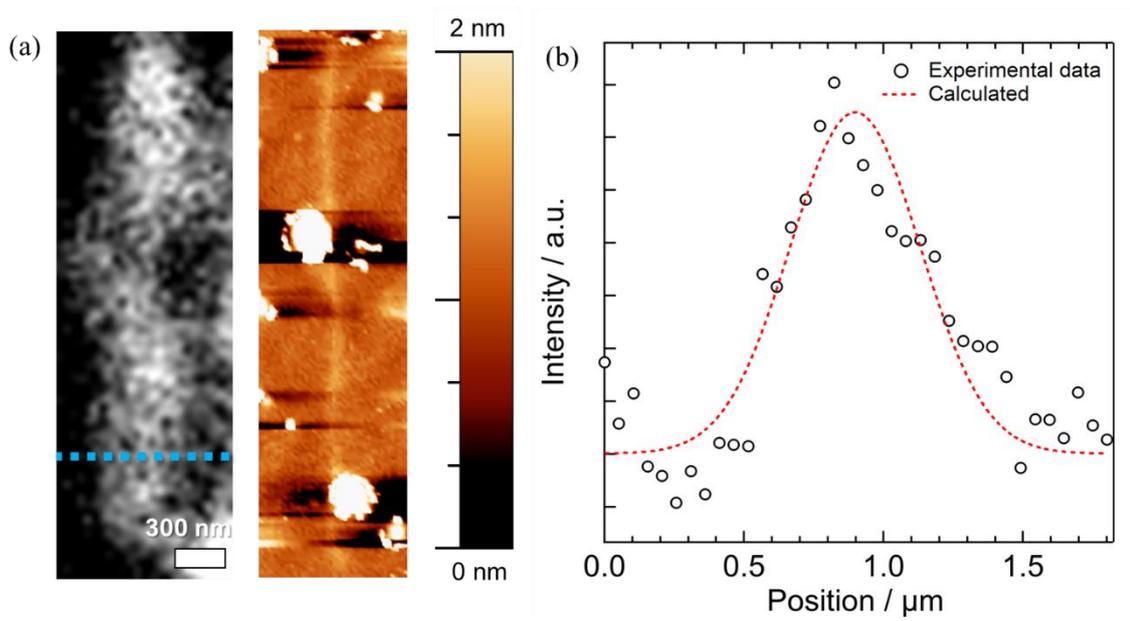

Figure 5 (a) Trion PL image and the corresponding AFM of hBN/MoSe$_2$/hBN with a single CNT back gate electrode. (b) Line profile along the blue dotted line shown in (a). Profile calculated through convolution of Gaussian and Lorentzian functions is also shown as the red dashed line.